\documentclass[twocolumn,amsmath,amssymb,prb]{revtex4}
\usepackage{latexsym}
\usepackage{graphicx}
\usepackage{color}
\begin{document}
\title{Transport Properties of Near Surface InAs Two-dimensional Heterostructures}

\author{Kaushini~S.~Wickramasinghe$^{1,2}$}
\author{William~Mayer$^{1}$}
\author{Joseph~Yuan$^{1}$}
\author{Tri~Nguyen$^{3}$}
\author{Lucy~Jiao$^{1}$}

\author{Vladimir~Manucharyan$^{2}$}
\author{Javad~Shabani$^{1}$}
\affiliation{$^{1}$Center for Quantum Phenomena, Department of Physics, New York University, NY 10003, USA
\\
$^{2}$Department of Physics, University of Maryland, College Park, MD 20742, USA
\\
$^{3}$Department of Physics, City College of City University of New York, New York City, NY 10031, USA}

\date{\today}
\begin{abstract}

Two-dimensional electron systems (2DESs) confined to the surface of narrowband semiconductors have attracted great interest since they can easily integrate with superconductivity (or ferromagnetism) enabling new possibilities in hybrid device architectures and study of exotic states in proximity of superconductors. In this work, we study indium arsenide heterostructures where combination of clean interface with superconductivity, high mobility and spin-orbit coupling can be achieved. The weak antilocalization measurements indicate presence of strong spin-orbit coupling at high densities. We study the magnetotransport as a function of top barrier and density and report clear observation of integer quantum Hall states. We report improved electron mobility reaching up to 44,000 cm$^{2}$/Vs in undoped heterstructures and well developed integer quantum Hall states starting as low as 2.5~T.

\end{abstract}
\maketitle

Epitaxial two-dimensional heterostructures containing InAs layers are hypothesized to be suitable systems for spintronics applications \cite{ZuticRevModPhys}.  Spintronics favors materials like InAs with strong spin-orbit interaction (SOC) and large g-factor.  Recently, two-dimensional electron systems (2DESs) confined to surface InAs layers have become the focus of renewed theoretical and experimental attention partly because of their potential applications in superconducting quantum computation \cite{Maxim17, Karl17} and realization of topological states of matter \cite{Aliceareview, LutchynReview}. A key feature to these applications is controlled proximity effect with superconductors \cite{Sau12, Cole15}. InAs is well known to be among a few materials which can interface well with metals and superconductors forming ohmic contacts unlike well known Si and GaAs materials where the interface forms a Schottky barrier. In the past few decades, studies have been focused on high mobility 2DES where the quantum well is placed tens or hundreds of nanometers from the surface \cite{Nitta92, Kroemer_1994, Takayanagi95, Mur96, Heida98, Richter_1999, Hatke,Tschirky17,Shayegan17}. To make proximity devices in these heterostructures, contacts have to be made after etching the top barriers with in-situ ion cleaning. The fabrication difficulties have limited fabricating and testing of more complicated devices \cite{Lehnert} since the interface quality/characteristics are proven difficult to finely tune. With the renewed interest in the field of topological superconductivity near surface quantum well structures with epitaxial superconducting contacts were developed \cite{Shabani2016,Henri17, Fabrizio17}. 


\begin{figure}[htp]
\centering

\includegraphics[scale=0.455]{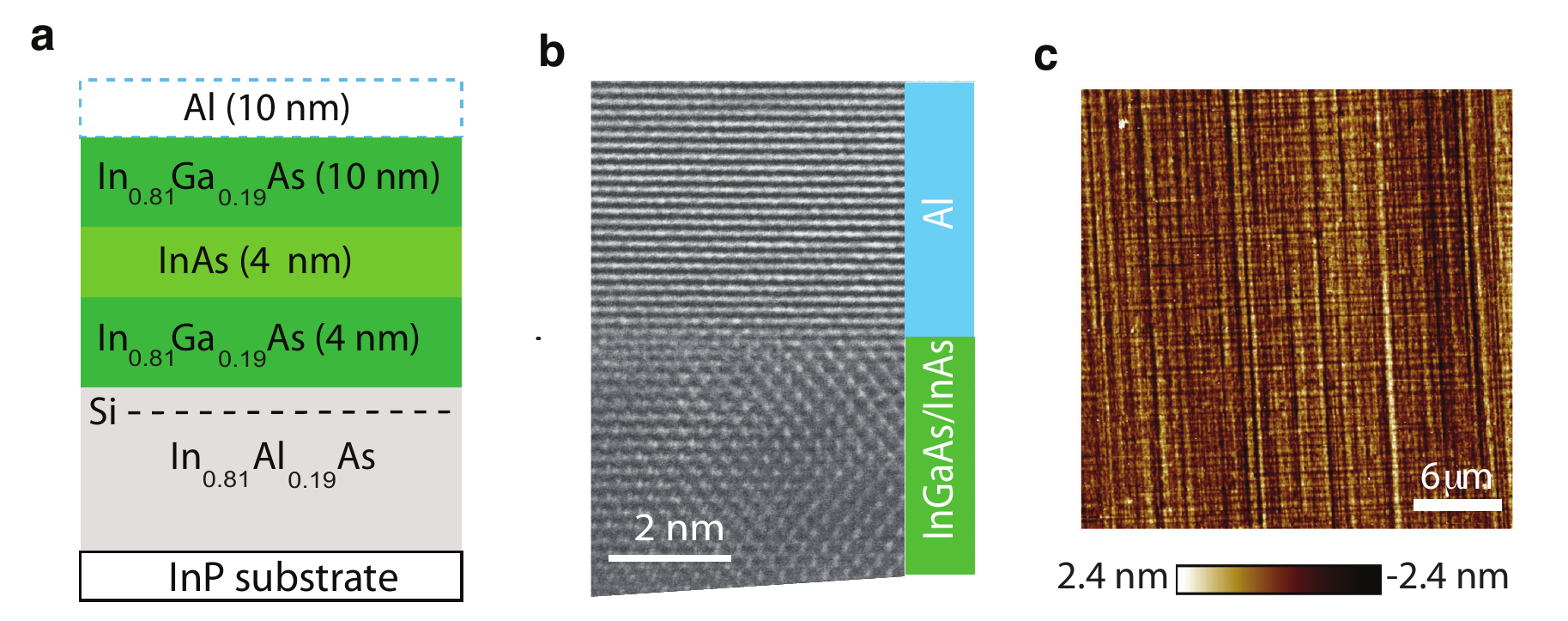}
\caption{(Color online) (a) General structure of hybrid Al-InAs surface quantum well.  (b) High-resolution transmission microscopy image showing that the Al forms a sharp and uniform interface to the InGaAs layer.  (c) Atomic force microscopy image shows the surface roughness of about 0.8 nm.}
\label{structure}
\end{figure}

The InAs-Al structure is well suited for the search of the so-called Majorana zero modes, where InAs has a strong SOC  and can be well interfaced with Al  \cite{Das12,Deng16}.  In order to probe and establish their unique statistics and eventually move beyond demonstrations of zero bias signatures to braiding \cite{AliceaNatPhy11, HalperinPRB12} and larger-scale Majorana networks \cite{AliceaBraiding,Matos2017}, it is likely that a top-down patterning approach will be needed.  Growth of large-area 2D Superconductor-Semiconductor systems through Molecular Beam Epitaxy (MBE) can provide the basis for such an approach. Josephson Junctions have been studied \cite{Shabani2016} and signatures of Majorana fermions have been explored \cite{Henri17,Fabrizio17} on such 2D hybrid InAs-Al platforms.  The question remains how well one can control the 2DES at the surface. There are key characteristics that should be considered when designing a surface 2DES to study the proximity effect using Josephson devices. The coupling of the 2DES with the surface is one of the important factors and achieved via the InGaAs insertion layer; the quality of the coupling depends on the composition and the thickness of this layer. A good carrier mobility should be achieved for an optimized insertion layer. Gate tunable mobility and density of carriers in a wide range and a strong SOC are also required. Near surface 2DESs are not only useful for tuning proximity effect but also advantageous in engineering new platforms to host other exotic particles. For example, if the 2DESs exhibit integer and fractional quantum Hall states one might be able to engineer a system to host parafermion quasiparticle \cite{Meng14,Clarke13}. Our goal is to develop InAs surface 2DES which is capable of demonstrating all the above characteristics to couple with a superconductor to tune the proximity effect using Josephson devices.

In this work we first study the magnetotransport properties of near surface InAs quantum wells and for the first time observe well developed integer quantum Hall states. We find that there is an optimum top barrier thickness that preserves reasonable mobility and wavefunction overlap at the surface. Also spin-orbit interaction measured by weak antilocalization (WAL) analysis shows a substantial strength only at higher carrier densities. The optimized structure provides a semiconducting platform with lower disorder and higher mobility than previously reported in the literature \cite{Shabani2016, kjaergaard2016transparent}. We find replacing InGaAs by InAlAs results in improvement of mobility for comparable thicknesses but the wavefunction overlap near the surface rapidly vanishes. In the last part of the paper, we discuss the limiting factors of the mobility and the possibility of further improving the quality of the structures but this comes at a lower density where we do not observe signatures of SOC.


%
\begin{figure}[htp]
\centering
\includegraphics[scale=0.545]{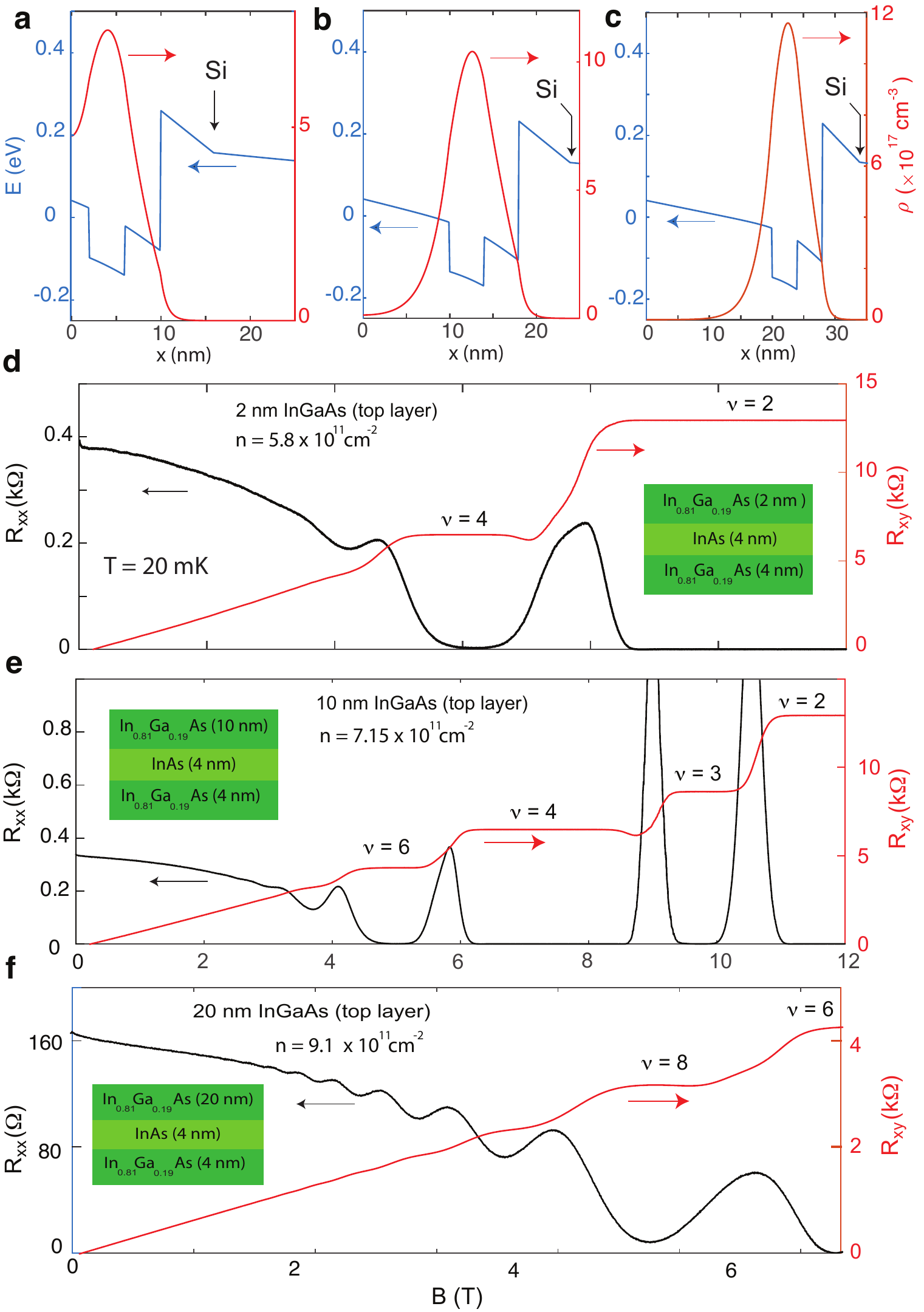}
\caption{(Color online) Self consistent Poisson charge distribution calculations for top layers of (a) 2 nm InGaAs, (b) 10 nm InGaAs and (c) 20 nm InGaAs.  Magnetotransport and layer structure of InAs quantum wells are shown for top layers of (d) 2 nm InGaAs, (e) 10 nm InGaAs and (f) 20 nm InGaAs.}
\label{QHplots}
\end{figure}


\begin{figure}[htp]
\centering
\includegraphics[scale=0.84]{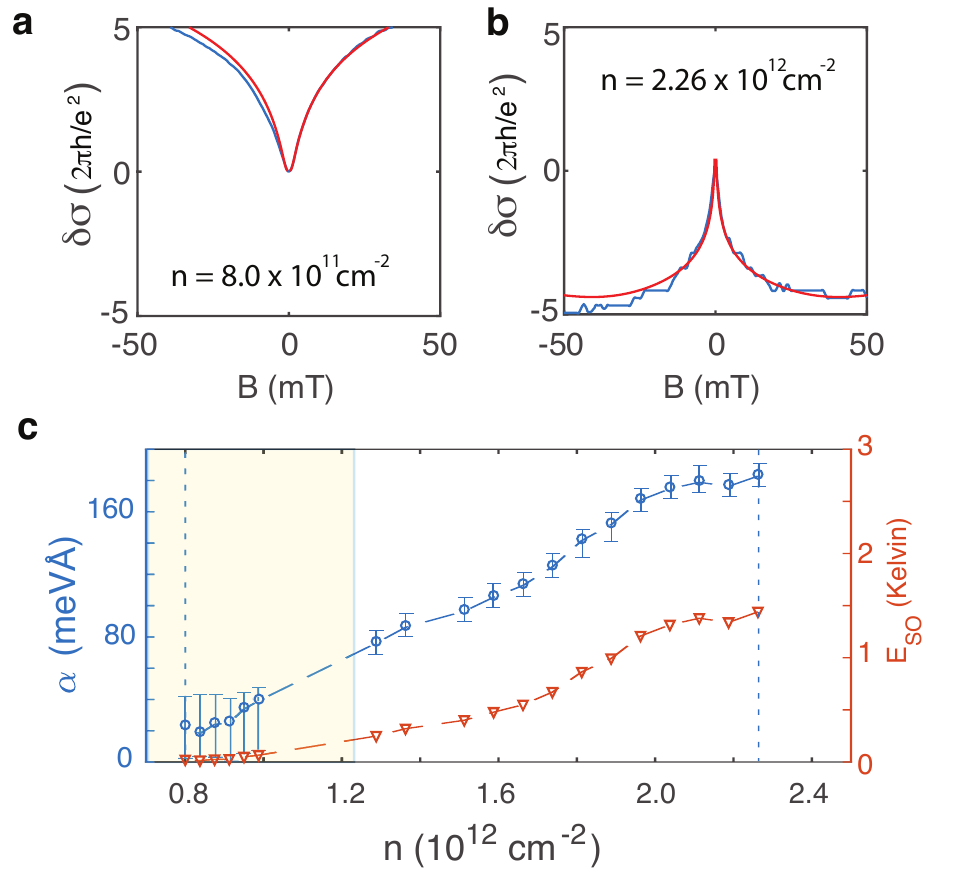}
\caption{(Color online) Low magnetic field response of InAs 2DES for densities:  (a) $n=8.0 \times 10^{11}$ cm$^{-2}$ (b) $2.26 \times 10^{12}$ cm$^{-2}$. The fit using ILP model is shown in red, see text. (c) Rashba SOC parameter, $\alpha$ (left axis) and corresponding spin-orbit gap $E_{SO}$ right axis) as a function of density. Dashed lines correspond to densities in part (a and b). Shaded regions in parts (c) denote $l_{SO} \geq l_{\phi}$ where the fits are no longer valid for extracting spin-orbit parameters.}
\label{WAL}
\end{figure}

The samples were grown on semi-insulating InP (100) substrates in a modified Gen II MBE system. The step graded buffer layer, In$_{x}$Al$_{1-x}$As, is grown at low temperature to minimize dislocations forming due to the lattice mismatch between the active region and the InP substrate \cite{Wallart05, ShabaniAPL2014, ShabaniMIT}.  The quantum well consists of a 4 nm layer of InAs grown on a 4 nm layer In$_{0.81}$Ga$_{0.19}$As. This layer helps to recover the InAlAs buffer roughness. We do not observe any improvement in transport mobility when the thickness of this layer is thicker than 4 nm consistent with earlier studies \cite{Wallart05}. A delta-doped Si layer of $\sim 7.5 \times 10^{11}$ cm$^{-2}$ is place 6 nm below the lower 4 nm In$_{0.81}$Ga$_{0.19}$As layer. A top layer, typically 10 nm of In$_{0.81}$Ga$_{0.19}$As, is grown on InAs strained quantum well as shown in Fig.~\ref{structure}a. After the quantum well is grown, the substrate is cooled to promote the growth of epitaxial Al (111)  \cite{Shabani2016}.  Figure \ref{structure}b shows a high-resolution transmission electron microscope (TEM) image of this interface between Al and In$_{0.81}$Ga$_{0.19}$As, with atomic planes of both crystals clearly visible.  All samples reported in this paper have had Al thin films grown and  thin films of Al were selectively removed, using Al etchant Transene solution type D, for transport studies of the InAs quantum well. The surface roughness of samples does not change with or without Al thin films confirmed by atomic force microscope images. Figure \ref{structure}c shows a surface topography of a sample with InGaAs top layer on a 34 $\mu$m by 34$\mu$m square.

The Fermi level pinning at semiconductor surfaces has been the subject of numerous theoretical and experimental studies. In most semiconductors, such as GaAs, the Fermi level is pinned inside the band gap \cite{MeadPR64}. It is well known that the surface states in case of InAs can result in a two dimensional electron system \cite{TsuiPRL70}. The electron accumulated is due to pinning of the Fermi level above the conduction band minimum. The position of the pinning level depends on the crystal direction and the surface treatments \cite{Sulfur}. Experiments on In$_{x}$Ga$_{1-x}$As predict Schottky barrier height becomes negative, exhibiting an ohmic behavior, for $x > 0.85$ \cite{KajiyamaAPL73}. Low temperature transport properties of surface accumulation InAs structures are dominated by surface scattering (low mobility, limited to a 1000-3000 cm$^2$/Vs) and magnetotransport shows very weak signatures of Shubnikov-de Haas oscillations. Adding a top layer to pure InAs quantum well resolves these issues. Figure ~\ref{QHplots}(d,e,f) shows the longitudinal and Hall resistance traces of van der Pauw samples with 2, 10 and 20 nm InGaAs top layers. The corresponding  charge distributions are shown in Figure 2(a,b,c). All samples are prepared similarly and measurements are performed at 20~mK. The electron effective electron mass of InGaAs at $x$ = 0.81 is $m_{e}$ = 0.03$m_{0}$, where $m_{0}$ is bare mass. This is comparable to electron effective mass of InAs which results in a quantum well inside a quantum well structure, supporting confinement of wavefunction in both layers. However, wavefunction in the quantum well slowly decays towards the surface \cite{Shabani2016}.  Figure \ref{QHplots}d shows the magnetotransport data of a sample with 2~nm InGaAs top layer. As evidenced by charge distribution calculations in Fig.~\ref{QHplots}a, there is a strong overlap with the surface and suitable for strong coupling of the semiconductor to superconductors. The corresponding longitudinal and Hall data are shown in Fig.~\ref{QHplots}d. The onset of oscillations for the 2~nm InGaAs top layer sample is near 3 T ($\mu = 6,500$ cm$^{2}$/Vs). Increasing the thickness of InGaAs top layer to 10~nm improves the quality as shown in Fig.~\ref{QHplots}e with mobility increasing to $\mu = 14,400$ cm$^{2}$/Vs  and a small but finite overlap of the wavefunction at the surface. Increasing the top layer thickness to 20 nm lowers the visibility of quantum Hall states and slightly lowers the transport mobility to $\mu = 12,570$ cm$^{2}$/Vs. This quantum well has a vanishing wavefunction overlap near the surface as shown in the Figure~\ref{QHplots}(c). 

\begin{figure}[htp]
\centering
\includegraphics[scale=0.575]{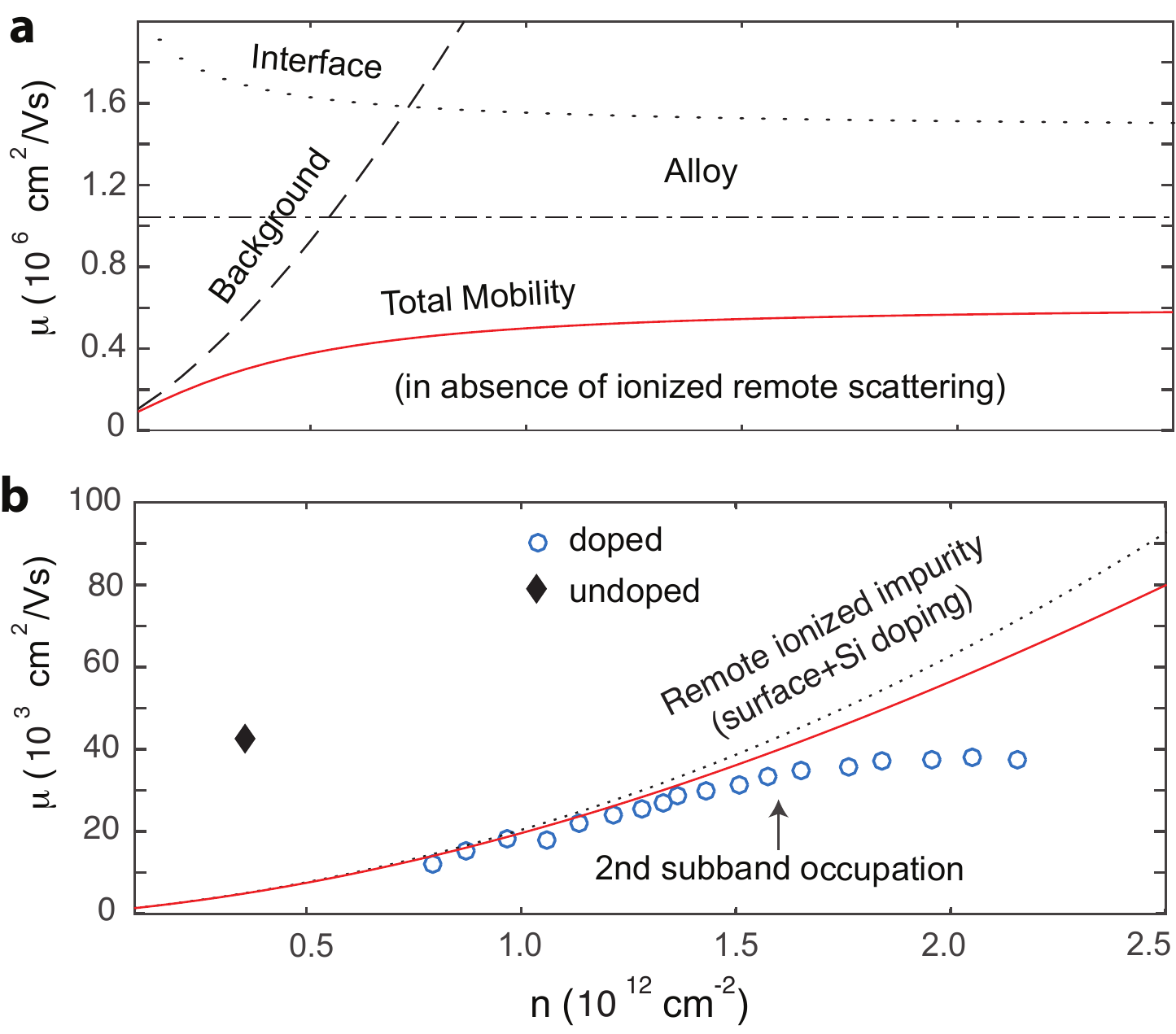}
\caption{(Color online) (a) Calculated mobilities from different scattering mechanisms as a function of doping electron density for buried quantum wells.  (b) Total theoretical mobility versus measured mobility for surface quantum wells. Solid red lines correspond to the total calculated mobilities where as the dash lines correspond to calculated mobilities due to individual scattering mechanisms.}
\label{DenCal}
\end{figure}

Engineering the coupling of the 2DES with surface is an important parameter in tuning the proximity effect. The goal is to keep the finite charge distribution at the surface in addition to highest electron mobility and quality of magnetotransport. We have studied several samples with various InGaAs top layer thicknesses and have found that a 10~nm top layer thickness yields the highest quality 2DES as well as good wavefunction overlap at the surface. Josephson devices fabricated on these structures exhibit $I_{c}R_{N} \sim 486 \mu V$. The magnetotransport with corresponding charge distribution are shown in Fig.~\ref{QHplots}(b,e). While electron density has not changed much, the quality has increased with onset of oscillations at 2.25 T and $\mu = 14,400$ cm$^{2}$/Vs. We also observe well developed filling factors $\nu$ = 3 and 6 which were not present in thinner top layers. Recent studies show that having a strong spin-orbit coupling may be compromised in strongly coupled limit, and somewhat weaker interface tunneling may be necessary for achieving optimal proximity \cite{Cole15} and estimated to be 10 nm in case of In$_{0.81}$Ga$_{0.19}$As \cite{Shabani2016}. For the rest of this paper we focus on properties of structures with 10~nm thick In$_{0.81}$Ga$_{0.19}$As top layer.

We next analyze the spin-orbit coupling using low field weak antilocalization measurements. These measurements are performed on gated Hall bars fabricated using optical photolithography and wet etch techniques.  Figure~\ref{WAL}(a,b) show the measured corrections to conductivity for two densities (data in blue). At $n=8.0 \times 10^{11}$ cm$^{-2}$, WAL signal shows absence of strong Rashba coupling while with increasing density, a peak appears and develops into a strong peak. Figure~\ref{WAL}(b) shows well developed WAL signal with a strong peak for the highest density that we probe, $n=2.26 \times 10^{12}$ cm$^{-2}$. The data is analyzed using the theory developed by Iordanski, Lyanda-Geller, and Pikus (ILP) for 2DESs \cite{ILP}. To reduce the number of free fitting parameters we fixed the value of cubic Dresselhaus SOC, $\gamma$ as the bulk value of InAs 26.9 eV $\AA^3$ calculated from the $\vec{k} \cdot \vec{p}$ theory \cite{Knap96}. Fitting $\delta \sigma (B) = \sigma(B) - \sigma(B=0)$ over the range $|B|<50$ mT at T = 20~mK yields the linear spin-orbit coupling parameter, $\alpha$ shown in Fig.~\ref{WAL}c. At high densities, Rashba SOC parameter reaches about 200 meV$\AA$ and spin-orbit energy gap, $E_{SO}$ reaches 1.5~K. Similarly, $l_{\phi}$ reaches to about 8~$\mu$m and $l_{SO}$ to 100 nm with a ratio of $l_{\phi}/l_{SO} \sim 80$. The dependence of $l_{so}$ and mean free path suggest that Dyakonov-Perel mechanism is dominant in our sample \cite{GrundlerPRL00,HerlingPRB17}.

%
%

\begin{figure}[tp]
\centering
\includegraphics[scale=0.45]{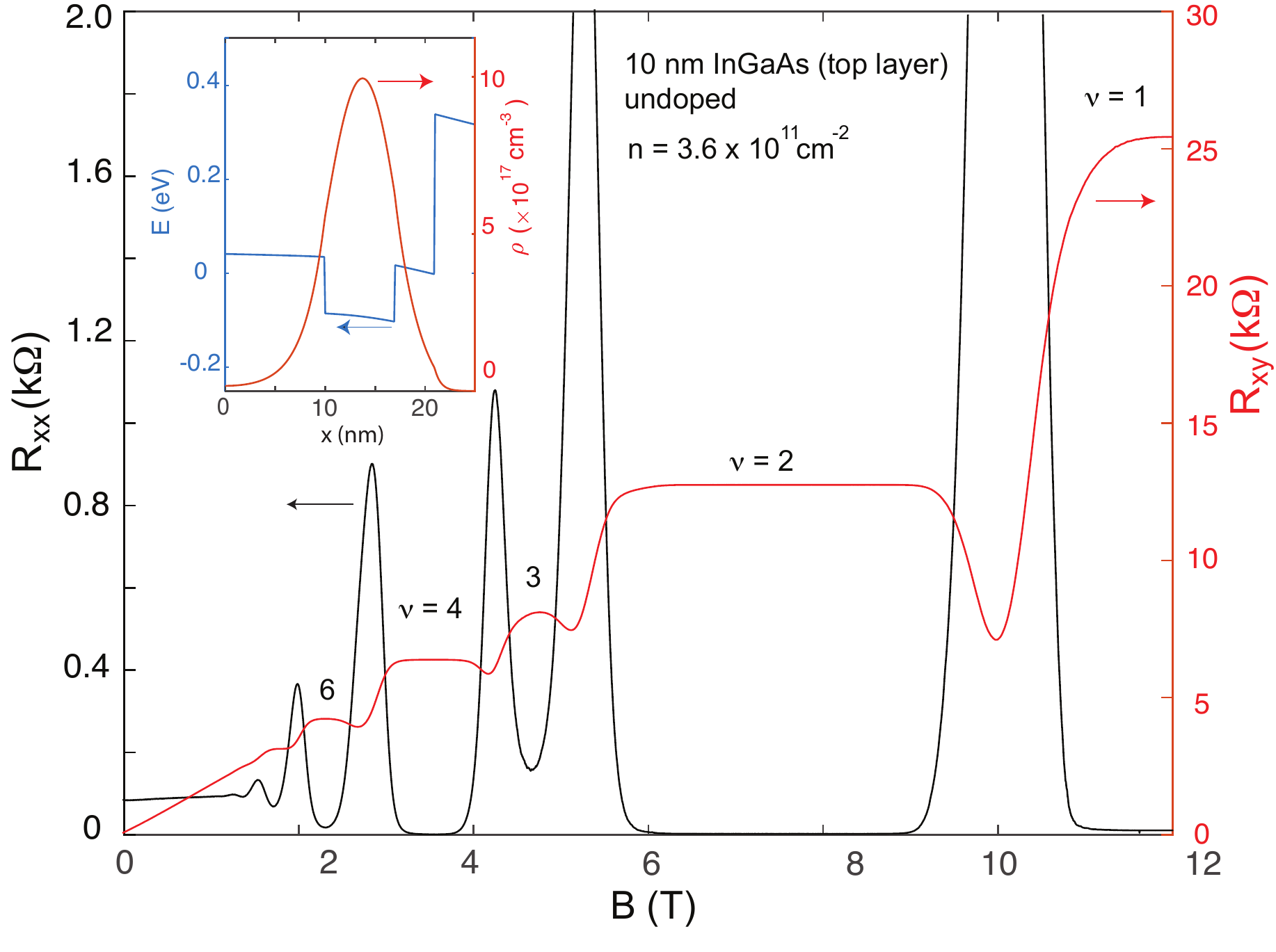}
\caption{(Color online) Magnetotransport for an undoped sample in the van der Pauw geometry with top layer of 10 nm  In$_{0.81}$Ga$_{0.19}$As.(Inset) Corresponding charge-distribution calculated for $n = 3.6 \times 10^{11}$ cm$^{-2}$.}
\label{139}
\end{figure}

The density dependence of the electron mobility can illuminate the various sources contributing to scattering. Fig.~\ref{DenCal}a shows calculated mobility due to scattering from rough interface, alloy scattering and background doping which are all relevant for deep structures (100 to 200 nm of InAlAs top layer) \cite{ShabaniMIT, ShabaniAPL2014}. The alloy scattering is highly sensitive to the ternary composition, $\mu \sim \frac{1}{x(1-x)}$; calculation is done following Ref.~\cite{AlloyPRB85}. The background doping, $N_{B}$, is deduced based on the observation that undoped deep structures conduct with density range in few $10^{11}$ cm$^{-2}$. Based on series of comparison between calculation and measurements we estimate $N_{B} \sim 10^{16} $cm$^{-3}$. The interface roughness with fluctuation height 0.8~nm adds a cut-off at $1.5 \times 10^{6}$ cm$^{2}$/Vs \cite{Shojaei16}. Using these values, the total mobility is calculated based on the Matthiessen's rule as shown in red in Fig.~\ref{DenCal}a. For electrons confined deeper in the structures (and not near surface), mobility is expected to increase with electron density and saturates around 600,000 cm$^{2}$/Vs similar to earlier studies \cite{ShabaniAPL2014}. However recent results in similar structures show higher mobilities can be achieved \cite{Hatke} on 120 nm deep structures which may suggest a discrepancy between experiments and alloy scattering model used by Ref.~\cite{AlloyPRB85}. In binary InAs quantum wells the electron mobility can even reach up to $\mu$ = 2.4 $\times 10^{6}$ and signatures of fractional quantum Hall states hav been observed \cite{Tschirky17,Shayegan17}. Note that the mobility in our structures are traded to achieve epitaxial superconducting contacts which can not readily happen in deeper 2DESs.

For surface quantum wells, the situation is different where surface scattering and remote ionized doping play major roles. The dependence of the mobility on density of our sample is shown in Fig.~\ref{DenCal}b as open blue circles. The electron mobility varies between 10,000 cm$^{2}$/Vs and 40,000 cm$^{2}$/Vs. The mobility starts to saturate near $1.6 \times 10^{12}$ cm$^{-2}$ close to second subband occupation (confirmed by calculation). The data can be fitted by $\mu \sim n^{\alpha}$, with $\alpha = 0.9$ in density range below $1.6 \times 10^{12}$ cm$^{-2}$.  The contribution of background doping is low in surface quantum wells as seen in Fig.~4a. The main scattering mechanism is surface and depends on details of scatter centers. For simplicity, we assume that the surface scattering contributions are similar to remote ionized impurities \cite{Gold87}. We first fit the 44,000 mobility data point from our undoped sample which results in $10^{11}$ cm$^{-2}$ ionized impurity.  We use this number and add intentional Si doping, $N_{d} \sim 7.5 \times 10^{11}$ cm$^{-2}$ and plot the combined results in Fig.~\ref{DenCal}b.  Solid red line corresponds to the total calculated mobility where as the dash line corresponds to calculated mobility due to remote ionized impurities (both surface and Si doping).
\begin{table}
  \caption{Summary of InAs surface quantum wells and their transport properties. Top layer material with indium composition of 0.81,  barrier thickness in nm: d, Si doping (cm$^{-2}$): N$_{d}$, Electron density (cm$^{-2}$): n, Mobility: $\mu$ , onset of Shubnikov-de Haas (Tesla): B$_{on}$.}
\begin{center}
\scalebox{0.9}{

    \begin{tabular}{|c|c|c|c|c|c|}
    \hline
    Barrier & d (nm) & N$_{d}$ (cm$^{-2}$)& n (10$^{11}$cm$^{-2}$)&$\mu$ (cm$^{2}t$/Vs)& B$_{on}$ (T)\\ \hline
    InAlAs & 5 & $10^{12}$  &  10 &  14,000 & 1.5  \\ \hline
    InAlAs & 2 & $10^{12}$  & 6.76 & 8,500 & 2  \\ \hline
    InGaAs &2 & $10^{12}$  & 5.8 & 6,500 & 3  \\ \hline
   InGaAs &10 & $10^{12}$  & 7.15 & 14,400 & 2.25  \\ \hline
    InGaAs &10 & -  & 3.6 & 44,000 & 1.2  \\ \hline   
    InGaAs &20 & $10^{12}$  & 10 & 12,570 & 1.3  \\ \hline

    \end{tabular}}
\vspace{-0.5cm}

\end{center}
\end{table}
We have also studied a van der Pauw sample without Si doping. Figure~\ref{139} shows the longitudinal and Hall data at T = 20~mK. As expected the electron density is reduced to $n = 3.6 \times 10^{11}$ cm$^{-2}$ but electron mobility is enhanced to 44,000 cm$^{2}$/Vs. Ref. \cite{JoonSue} reports online on arXiv mobility of $\mu$ = 54,000 cm$^2$/Vs  at a higher density of n = 8.7 $\times 10^{11}$. However they do not present magnetotransport data. In our sample, various integer quantum Hall states are clearly resolved at $\nu$ = 1, 2, 3, 4, 6, 8 with onset of oscillation at 1.2 T. We should note that our undoped structures initially did not conduct probably due to surface compensation. Originally during the growth, we would change the indium and aluminum cell temperatures to change the composition during step graded buffer. In all the structures we have presented in this paper, we have kept the Al cell temperature fixed while indium cell temperature has been adjusted during step graded buffer. We found that the optimum arsenic flux and growth temperatures are different and it is only in these structures that undoped surface quantum wells are conducting. The self consistent calculation estimates a carrier density of $2.16 \times 10^{11}$ cm$^{-2}$ from background doping and $\sim 1 \times 10^{11}$ cm$^{-2}$ from surface with total density of $3.16 \times 10^{11}$ cm$^{-2}$ which is close to the measured density. This suggests that in optimized structures the mobility can be increased without intentional doping. However we should note that in agreement with Fig.~3c we do not observe weak antilocalization signal at $n = 3.6 \times 10^{11}$ cm$^{-2}$.

In summary, we studied the scattering mechanism of InAs surface quantum wells with InGaAs top layer. In these structures, we find top layer thickness and ionized impurity scattering both affect the mobility. We find that in an undoped structure with a 10 nm top layer, the mobility increases up to 44000~cm$^2$/Vs and observe reduction of onset of oscillation in magnetotransport data down to 1.2 T with well developed integer states emerging at 2.5 T. We should also note that another choice for top layer is InAlAs. Transport properties are reported in Table I. 

Our work was supported by US Army research office and DARPA TEE. We thank Kasra Sardashti for fruitful discussions.

\bibliography{References_Shabani_Growth.bib}
\end{document}